\documentclass[preprint,prb,groupedaddress,floatfix]{revtex4}
\usepackage{graphicx,subfigure,amsmath,epsfig,amsfonts}
\usepackage{color}
\usepackage[sort&compress]{natbib}
\usepackage{subfigure}
\usepackage{rotating}
\usepackage{multirow}
\usepackage{bm}

\begin{document}

\title{Flow-dependent unfolding and refolding of an RNA by nonequilibrium umbrella sampling}

\author{Alex Dickson}
\affiliation{James Franck Institute, The University of Chicago, Chicago, IL 60637}

\author{Mark Maienshein-Cline}
\affiliation{James Franck Institute, The University of Chicago, Chicago, IL 60637}

\author{Allison Tovo-Dwyer}
\affiliation{James Franck Institute, The University of Chicago, Chicago, IL 60637}

\author{Jeff R. Hammond}
\affiliation{Leadership Computing Facility, Argonne National Laboratory, Argonne, IL 60439}

\author{Aaron R. Dinner}
\affiliation{James Franck Institute, The University of Chicago, Chicago, IL 60637;\\
correspondence to dinner@uchicago.edu}
\date{\today}

\begin{abstract}

Nonequilibrium experiments of single biomolecules such as force-induced unfolding reveal details about a few degrees of freedom of a complex system.
Molecular dynamics simulations can provide complementary information, 
but exploration of the space of possible configurations is often hindered by large barriers in phase space that separate metastable regions.
To solve this problem, enhanced sampling methods have been developed that divide a phase space into regions and integrate trajectory segments in each region.
These methods boost the probability of passage over barriers, and facilitate parallelization since integration of the trajectory segments does not require communication, aside from their initialization and termination.
Here we present a parallel version of an enhanced sampling method suitable for systems driven far from equilibrium:  nonequilibrium umbrella sampling (NEUS).
We apply this method to a coarse-grained model of a 262-nucleotide RNA molecule that unfolds and refolds in an explicit flow field modeled with stochastic rotation dynamics.
Using NEUS we are able to observe extremely rare unfolding events that have mean first passage times as long as 1.4 s ($3.4 \times 10^{13}$ dynamics steps).
We examine the unfolding process for a range of flow rates of the medium, and we describe two competing pathways in which different intramolecular contacts are broken.

\end{abstract}

\maketitle

\section{Introduction}

Nonequilibrium measurements on biological macromolecules, such as mechanical
force-induced unfolding \cite{Liphardt2001} and flow-based analogs
\cite{Lin2009}, have emerged as a powerful complement to equilibrium studies.
Indeed, it now possible to follow the evolution of distances through
fluorescence resonance energy transfer (FRET) simultaneously with forces
through optical traps \cite{Comstock2011}.   While these
measurements provide unprecedented experimental data on the stochastic dynamics
of individual molecules, they still only probe at most a few degrees of freedom
among many.  Molecular dynamics simulations, which provide complete information
about the positions of all participating particles subject to the assumptions
of the model, have proven to be a valuable tool for interpreting these
data \cite{Sotomayor2007}.  However, the time scales for conformational change
are often long compared with elementary fluctuations, which makes waiting for
the events of interest to occur spontaneously under conditions representative
of experimental ones prohibitively computationally costly.  To accelerate
convergence, many simulation studies employ unrealistically extreme
nonequilibrium conditions (see discussion in Hu \textit{et al.} \cite{Hu2006b}).

Alternatively, enhanced sampling methods can be used to improve exploration of
phase space and focus computational effort on low probability regions of
mechanistic importance, such as transition states.  The most widespread such
methods \cite{Frenkel2002,Dellago2009,E2010} rely on the fact that the
statistics of equilibrium systems are known {\it a priori}, which prevents the
applicability of such methods to nonequilibrium situations.   However, there
now exist methods that can enhance the sampling of low probability regions
without relying on equilibrium properties of the system
\cite{Allen2005,Allen2006poly,Allen2006maier,Allen2009_review,Warmflash2007,Dickson2009,Dickson2009_rate,Dickson2010_review,Huber1996,Zhang2007,Bhatt2010,Bhatt2010_semiatom}.
Although these methods differ in detail, the essential idea in all of them is
to harvest segments of unbiased dynamics trajectories such as to achieve
relatively uniform sampling of different regions of a space of physically
relevant degrees of freedom (order parameters).   The acceleration of
convergence follows from the fact that each trajectory segment need only
traverse a small portion of the space of order parameters, across which the
probability is relatively uniform.

We have been developing one such method:  nonequilibrium umbrella sampling
(NEUS) \cite{Warmflash2007,Dickson2009,Dickson2009_rate,Dickson2010_review}.
In this paper, we present a streamlined version of the algorithm with
improved convergence properties.  The most significant change is the explicit
association of a weight with each saved copy of the system, motivated by the
weighted ensemble method \cite{Huber1996,Zhang2007,Zhang2010,Bhatt2010,Bhatt2010_semiatom}.
The fact that many trajectory
segments are integrated independently makes the method highly parallelizable,
and we detail and implement a strategy that can provide excellent scaling to
large numbers of processors.  

We use the method to simulate partial unfolding and refolding of a
coarse-grained model of a 262-nucleotide RNA molecule in the presence of a flow
field.  Our interest in this system comes from single-molecule studies of FRET
between probes on the L18 loop and 3$'$ terminus of the catalytic domain of the
RNase P RNA from {\it Bacillus stearothermophilus}
\cite{Smith2008,Qu2008,Li2009}.  In these studies, the molecule was tethered in
a microfluidic channel to enable relatively rapid changes in magnesium ion
concentration, and this led to the question of whether flow contributed to the
dynamics observed \cite{Li2009}.  Here, we show that there are two competing
unfolding pathways, the likelihoods of which depend on the rate of flow of the
solution.  We compare these results with reversible unfolding simulations
(without a net flow).

\section{Methods}

\subsection{Algorithm}
\label{sec:alg}

As we show, the events of interest are on the time scale of milliseconds to seconds, 
while straightforward simulations of the coarse-grained model are limited to tens of microseconds.
Thus enhanced sampling is needed.  
Here, we describe the version of nonequilibrium umbrella sampling (NEUS)
\cite{Warmflash2007,Dickson2009,Dickson2009_rate,Dickson2010_review} used in
the present study.  To this end, we summarize the overall strategy, and then we
describe the phases of the simulation and parallelization;
differences from earlier versions of the algorithm and competing methods are
noted.

\subsubsection{Overall strategy}
\label{sec:ww}

The sampling is guided by a set of physically relevant variables (``order
parameters'').  Ideally, these order parameters will describe the slow
dynamics in the system, and the remaining degrees of freedom will relax
relatively fast.  In this work, we employ a single order-parameter that
quantifies the total number of intramolecular contacts (Section
\ref{sec:system}).  However, we explicitly separate the ``forward'' (unfolding)
and ``backward'' (refolding) transition path ensembles as in Dickson \textit{et
al.} \cite{Dickson2009_rate}.  This allows the sampling of the orthogonal
degrees of freedom to differ between the two ensembles (i.e., allows for
separate unfolding and refolding pathways), and it enables the calculation of
transition rates between basins.  

For the simulations, we divide the space of order parameters into regions,
which need not be uniform in size.  Each region contains one or more copies of the system
(walkers) 
that evolve independently according to the natural dynamics of the system, and we associate
with that copy a weight for contributing to averages.  When
a copy of the system attempts to leave its region, the configuration is saved
to a list of entry points for the neighboring region, along with the weight of
the copy.  When a neighboring list is full, the oldest saved configuration is
overwritten and its weight is distributed over the remaining points in the list
in a manner that does not affect their relative probabilities of being chosen.
The copy is then restarted from a saved configuration, $i$, which is chosen from one of
its region's lists with likelihood proportional to its weight ($w_i$).  The
weight of this point is then partitioned between the active and saved copies:
$\gamma w_i$ ($\gamma \in (0,1]$) is given to the active copy, and the rest,
$(1-\gamma) w_i$, remains associated with the saved entry point.  Note
that $\gamma = 1$ results in straightforward dynamics, or a single, continuous
trajectory.  Here we use $\gamma = 0.9$.  The incorporation of this feature in
the NEUS algorithm is motivated by the (equal) partitioning of the probability
when a trajectory branches in the weighted ensemble (WE) method
\cite{Huber1996}; it ensures conservation of the starting probability and
suppresses artificial amplification of particular trajectories.  As a result,
we are able to obtain converged results with only one set (lattice) of regions
in the extended space as opposed to two as in previous work
\cite{Warmflash2007,Dickson2009,Dickson2009_rate}.

\subsubsection{Initialization}

A common situation is that one is interested in studying a transition between
two or more states but one knows the configuration of the system in only one of
the stable states.  This situation applies here to the RNA-under-flow system, 
since we know the folded
configuration but not the most likely unfolded configurations.  Although in
principle one could start the simulation in each region using any configuration
consistent with the allowed order parameter values, in practice it is best to
start with a distribution of structures that is as consistent as possible with
the physically weighted dynamics to avoid introducing unnecessary errors that
take time to be corrected.  To this end, we progressively activate the regions
in a manner similar to Forward Flux Sampling (FFS) \cite{Allen2009_review} as follows.

We start by running an unconstrained simulation that is initialized in the known stable
configuration.  During this simulation, we record the configuration each time
the system crosses a boundary of a region but do not reset the configuration.
These configurations serve as the initial entry (i.e., resetting) points
for the regions visited, and all such configurations are assigned equal weight.
Following the unconstrained simulation, we begin the umbrella sampling
simulation starting from saved entry points in each region that has at least
one such point, employing and updating the copy weights as described above.
Regions that were not visited previously are activated once entry points for
them are obtained.  As the simulations proceed, regions of lower and lower
probability are activated by their neighbors, and trajectories emerge from the
original stable state.  Once all the regions are activated, we are able
to concurrently sample the entire order parameter space of interest, using only
points that resulted directly from the starting distribution.  

In the present study, the progressive initialization of regions accounts for
about 2\% of the total simulation time.  The sampling procedure employed here
further differs from FFS in that it does not explicitly require a notion of
forward progress and thus can be used with sampling regions that are defined by
an arbitrary number of order parameters.  By the same token, trajectories are
terminated when they cross any boundary, not only a forward one.   This
distinction is of practical importance when the dynamics do not lead rapidly
back to the starting basin (see Dickson \textit{et al.}
\cite{Dickson2010_review} for further discussion).

\subsubsection{Weight redistribution}

The algorithm as described is in principle complete.  Indeed, it is very
similar to the WE method except that (i) it permits strict control of the
number of copies in a region (including only limiting it to one) and (ii)
differs in the details of weight partitioning when resetting (branching) and
redistributing when overwriting (pruning).  However, the transfer of
weight between regions of high probability can be very slow when the weight 
must pass through a bottleneck 
region of low probability.  This is because
a very large number of low probability walkers are required to add up to a
significant change of weight in a high probability region.  This convergence
issue arises despite the fact that the time for initial exploration of the space 
decreases with increases in the number of regions, as in any umbrella sampling
procedure \cite{Chandler1987,Frenkel2002}.

To accelerate convergence after the initialization phase, we periodically use
the interface-to-interface crossing statistics to predict statistical weights
for each region ($\{W_i\}$), and scale the weights of the entry points in
each region, $i$, such that their sum is equal to $W_i$.  Here, the weights are
obtained from a modified version of the scheme in Vanden-Eijnden and Venturoli
\cite{Vandeneijnden2009_twist}, where the total flux into a region is set equal
to the total flux out of a region.  To this end, we accumulate a transition
matrix, $\mathbb{T}$:  each off-diagonal element $t_{ij}$ is the number of
transitions observed from region $j$ to region $i$ in the last weight update period, 
and each diagonal element
$t_{ii} = -\sum_{j} t_{ji}$.  We then solve the equation $\mathbb{T}{\bf
W}={\bf 0}$ for the weight vector ${\bf W}$ by using singular value
decomposition to compute the nullspace of $\mathbb{T}$, which is the single
nontrivial solution ${\bf W}$.   Here, we perform this operation periodically
throughout the simulation, as in previous NEUS studies
\cite{Dickson2009_rate,Vandeneijnden2009_twist}; this contrasts with the study
by Bhatt \textit{et al}. \cite{Bhatt2010} in which a single such step is used
to pre-condition the simulation and then flux balance is used to check
convergence.

\subsubsection{Parallelization}

The simulations of the copies of the system require only limited communication.
As such, NEUS and methods like it lend themselves well to parallelization.
However, we find that they benefit from careful implementation on high
performance computers.  All simulations for the present study are run on
parallel architectures using the Global Arrays toolkit \cite{Nieplocha1996},
which implements a global address space programming model in which processes
can access remote data using one-sided communication.  
One-sided communication is particularly useful in this case, since the timing
of boundary crossing events is not predictable.  The global address space
also enables one to distribute the storage of a large set of region
entry points across the memory of many compute nodes.  
The entry points for each region, the
region weights, boundary crossing statistics and sampling histogram data are
all stored as global arrays.  These arrays can be
modified by any process using ``put'' functions and ``get'' functions, where
locks are used to enable atomic updates of global data (modifications of
the entry point lists, for instance) that prevent processes from
concurrently accessing the same region of a global array.

Although the dynamics of the copies are simulated essentially without
communication once they are initialized, NEUS still periodically requires some
collective operations, such as weight updates, and the computation of rates and
probability distributions.  To allow for such operations, we break down the
simulation into ``cycles'' of computation, at the end of which all processes
are synchronized.  Within the cycles, the work is distributed among the
processes as follows.  When a process is finished running a trajectory segment,
it queries how many steps have been run in each region $k$ so far this cycle
($N_k$), and it uses the results to decide in which region to run the next
trajectory segment.  Specifically, it chooses to start a trajectory in
region $j$ with probability
\begin{equation}
P_j = \frac{N_{\text{steps}}-N_j}{\sum_{k}(N_{\text{steps}}-N_k)},
\end{equation}
where $N_\text{steps}$ is the number of steps to be run in each region per
cycle.  A trajectory is run until either the counter in its region reaches
$N_\text{steps}$ (upon which the current configuration of the system is saved
to the entry point list as a simple means of maintaining it), and a computational 
cycle ends when all counters reach $N_\text{steps}$.


\subsection{Model}
\label{sec:system}

The system is a model of the catalytic domain of RNase P RNA from
\textit{Bacillus stearothermophilus}.  To make the simulations tractable, we
use a coarse-grained representation that averages over the atomic structure and
dynamics, while taking into account the secondary and tertiary interactions
that stablize the native state: the self-organized polymer (SOP) model
\cite{Hyeon2007}.  In the SOP model, each nucleotide of the RNA is treated as a
bead, and the beads interact through potentials that depend on the known native
structure.  The potential defining the model is the sum of a finitely
extensible nonlinear elastic (FENE) potential that connects adjacent beads
\cite{Kremer1990} ($V_\text{FENE}$); a Lennard-Jones attraction between beads that
has a minimum at the native structure distance ($V^{\text{att}}_{\text{nb}}$);
pairwise non-bonded repulsions scaling as $r^{-6}$, which locally straighten
the chain and mimic steric repulsions between nucleotides
($V^{\text{rep}}_{\text{nb}}$); and a Weeks-Chandler-Andersen \cite{Weeks1971} (WCA)
repulsion between each bead and the wall at $y=0$ ($V^{\text{wall}}$).  The
total potential function is
\begin{equation}
V_{\text{T}} = V_{\text{FENE}} + V^{\text{att}}_{\text{nb}} + V^{\text{rep}}_{\text{nb}} + V_{\text{wall}}\nonumber\\
\end{equation}
with
\begin{align}
  \label{eq:V}
V_{\text{FENE}}&= -\sum_{i=1}^{N-1} \frac{k}{2} R_0^2 \log\left(1-\frac{(r_{i,i+1}-r^0_{i,i+1})^2}{R_0^2}\right) \nonumber \\
V^{\text{att}}_{\text{nb}}&= \sum_{i=1}^{N-3}\sum_{j=i+3}^N \epsilon_h \left[\left(\frac{r_{ij}^0}{r_{ij}}\right)^{12}-2\left(\frac{r_{ij}^0}{r_{ij}}\right)^6\right]\Delta_{ij} \nonumber \\
V^{\text{rep}}_{\text{nb}}&=\sum_{i=1}^{N-2} \epsilon_l \left(\frac{\sigma^*}{r_{i,i+2}}\right)^6 + \sum_{i=1}^{N-3} \sum_{j=i+3}^N \epsilon_l \left(\frac{\sigma}{r_{ij}}\right)^6(1-\Delta_{ij}) \nonumber \\
V_{\text{wall}} &= \sum_{i=1}^{N} H(2^{1/6}\sigma_{\text{WCA}}-y_i) \times 4 \epsilon_l \left[ \left( \frac{\sigma_{\text{WCA}}}{y_i}\right)^{12} - \left( \frac{\sigma_{\text{WCA}}}{y_i}\right)^{6} \right],
\end{align}
where $r_{ij}$ is the distance between residues $i$ and $j$, and $r^0_{ij}$ is
their distance in the native structure.  We set the parameters in (\ref{eq:V})
to those in Hyeon and Thirumalai \cite{Hyeon2007}, namely 
$R_0=0.2$ nm, $\epsilon_h = 0.7$ kcal/mol, and $\epsilon_l= 1.0$ kcal/mol; we set 
$\sigma = 7$ {\AA} to ensure noncrossing of the chain,
and we set $\sigma^* = 3.5$ {\AA} to prevent the flattening of helical structures.  
In $V_{\text{wall}}$, $\sigma_{\text{WCA}} = 2$ {\AA}, and $H(x)$ is a Heaviside
function equal to $0$ for $x < 0$ and $1$ for $x > 0$.  The equation of motion
for the polymer is integrated with the Velocity-Verlet algorithm with time step
$\delta t = 40$ fs.

\begin{figure}
\begin{center}
\includegraphics[width=6 in]{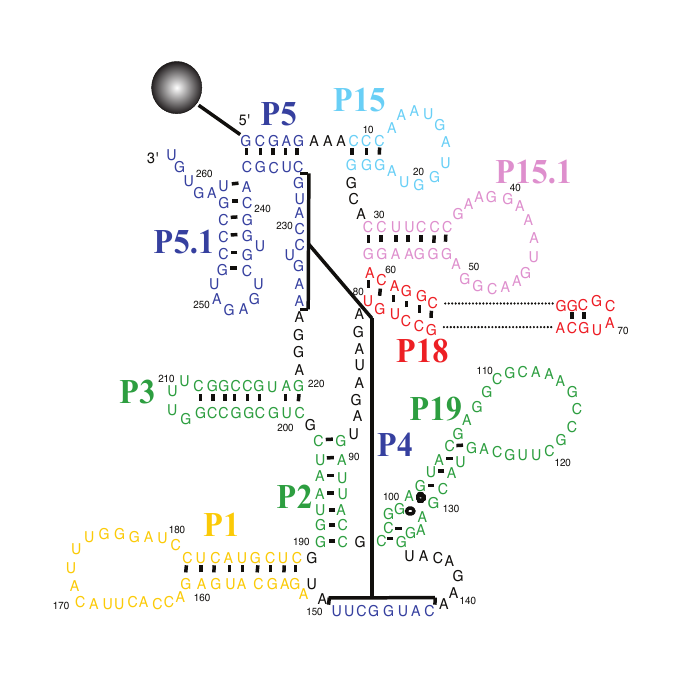}
\caption{\label{fig:sec} 
Secondary structure of the RNA molecule.  In the simulation, the 5$'$ end of the molecule is attached to a tether (black sphere), that prevents the molecule from moving along with the flow.  The index of every tenth residue is shown.
}
\end{center}
\end{figure}

The native, folded structure was constructed from the crystal structure for the full 
RNase P RNA \cite{Kazantsev2005}.  The coordinates of the catalytic domain (262 residues) were isolated 
from the full structure (417 residues), and coarse-graining into beads was carried 
out by replacing the coordinates of each residue with its center of mass. Unstructured 
residues, which did not have crystal structure coordinates (in Figure \ref{fig:sec}, residues 161-181 
in P1, 15-20 in P15, 64-73 in P18, and 106-125 in P19) were added by introducing the 
appropriate number of beads into the sequence, separated by the average bead-bead 
distance (about 5 {\AA}); these unstructured residues have no contacts. The structure 
was allowed to relax to its minimum energy by integrating without a random force so that 
the added unstructured residues form simple loops.
Using this structure, we consider a native contact to exist ($\Delta_{ij} = 1$) 
between all pairs of residues
$i$ and $j$ with $|i-j|>2$ and distance less than $R_C =
1.4$ nm in the native structure; for all other pairs $\Delta_{ij} = 0$.

The solvent in the simulation is modeled using the stochastic rotation dynamics
method \cite{Malevanets1999,Ihle2001,Lamura2001,Allahyarov2002}, in which the
solvent is represented by a large number of infinitesimal particles that are
grouped into cubic ``interaction cells''.  Each step of the algorithm comprises 
two parts: (1) free streaming, in which the position of particle $i$ (${\bf
r}_i$) is updated according to ${\bf r}_i(t+\Delta t)= {\bf r}_i(t) + {\bf
v}_i(t)\Delta t$ where ${\bf v}_i$ is the velocity at time $t$ and $\Delta t =
150 \delta t$ is the solvent time step and (2) ``collision'', in which ${\bf
v}_i(t+\Delta t)= {\bf v}_{\rm cell}(t) + \Omega[{\bf v}_i(t) - {\bf v}_{\rm
cell}]$ where ${\bf v}_{\rm cell}$ is the average velocity of particles in the
cell containing $i$, and $\Omega$ is a stochastic rotation matrix which rotates
vectors around a random axis
by $\pm\alpha$, a fixed angle, with equal likelihood.
Here we use $\alpha =
0.243\pi$, which in combination with the other parameters used here for the
solvent, gives a viscosity of $0.8$ g/m/s, which is approximately the viscosity
of liquid water at our simulation temperature (300 K).  The viscosity was
calculated using Equations 10 and 14 of Kikuchi \textit{et al.}\cite{Kikuchi2003}.

We allow the solvent to influence the RNA by including the polymer beads in the collisions,
as in Webster and Yeomans\cite{Webster2005}.  This is
done using 
\begin{equation} {\bf v}_{\rm cell}(t) = \frac{\sum_{\rm cell} m{\bf
v}_i(t) + \sum_{\rm cell} M{\bf V}_i(t)}{N_{\rm cell}^{\rm solv}m + N_{\rm
cell}^{\rm poly}M} 
\end{equation} 
where $m = 32$ amu is the mass of the solvent particles (chosen to make a
solvent mass density of 1 g/mL), and $M = 300$ amu is the mass of the residues,
compared with a range in mass for RNA nucleotides of 320 to 360 amu.  ${\bf
V}_i(t)$ is the velocity vector for residue $i$, and the sum is over all
particles in the cell.  

\begin{figure}
\begin{center}
\includegraphics[width=6 in]{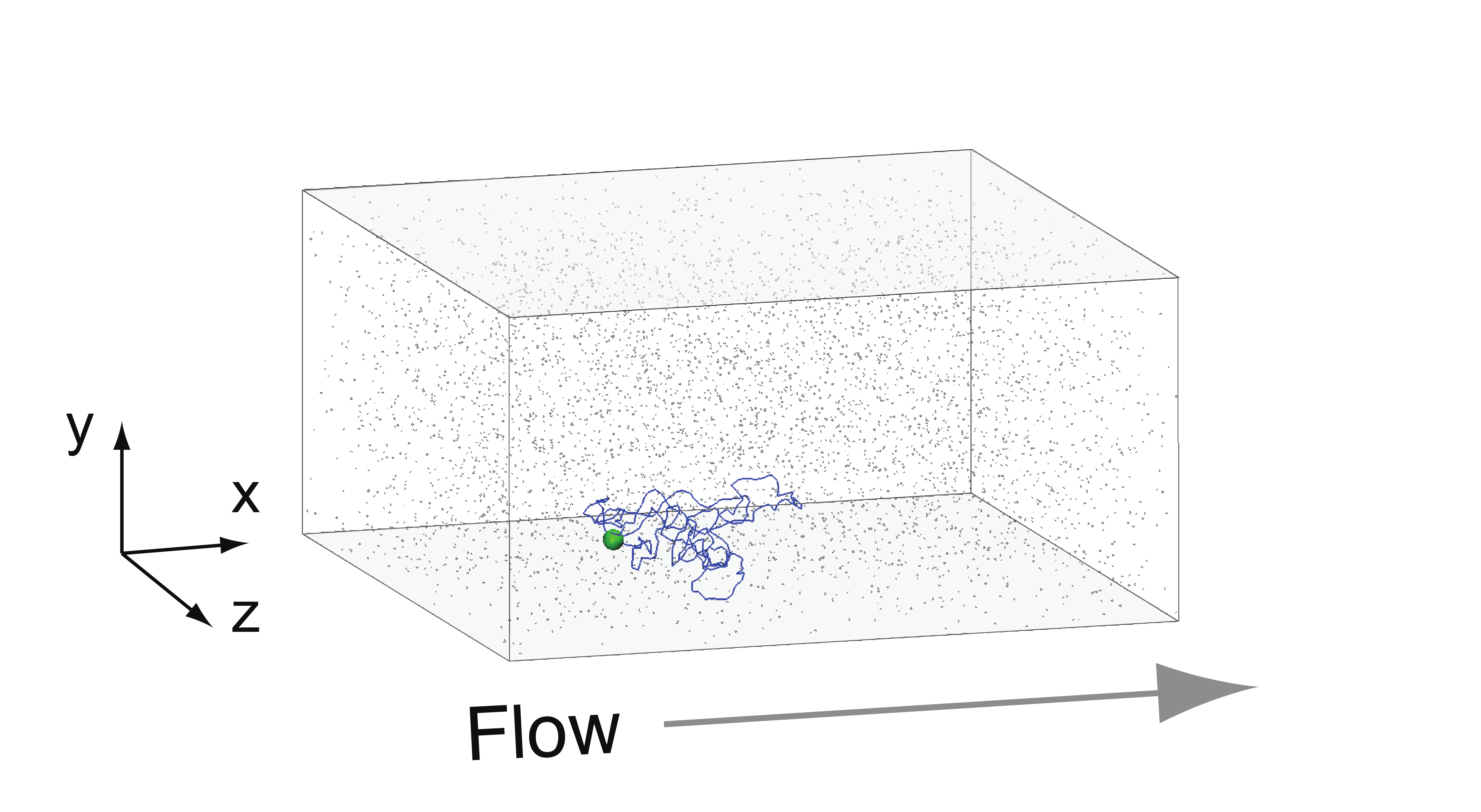}
\caption{\label{fig:flowbox} 
The simulation cell.  The boundaries at $y=0$ and $y=L_y$ have reflective boundary conditions, while the others are periodic.  The tether point is shown as a large green sphere, and the RNA molecule is in blue.  $5000$ of the $503200$ solvent molecules are shown here.  A flow is induced in the positive $x$-direction by applying a constant acceleration to the solvent particles, which in turn causes extension of the RNA molecule in that direction.
}
\end{center}
\end{figure}

We use periodic boundary conditions in the $x$ and $z$ directions, reflective
walls at $y=0$ and $y=L_y$, and drive the solvent to flow in the positive $x$
direction (Figure \ref{fig:flowbox}).  The dimensions of the box are $L_x = L_z
= 384$ {\AA} and $L_y = 192$ \AA. The interaction cells are cubic with side
length 8 \AA, which was chosen to be comparable with the average distance
traveled by a solvent particle in a time $\Delta t$.  Following previous work,
we shift the lattice periodically to avoid artifacts \cite{Ihle2001} and employ
the generalized bounce back rule for partially filled cells along the $y=0$ and
$y=L_y$ edges \cite{Lamura2001}.  
An extra FENE interaction is added between the 5$'$
terminus and the tether point, located at (120 \AA, 25 \AA, 192 \AA) to prevent
the molecule from moving along with the flow.  

\begin{figure}
\begin{center}
\includegraphics[width=3.375 in]{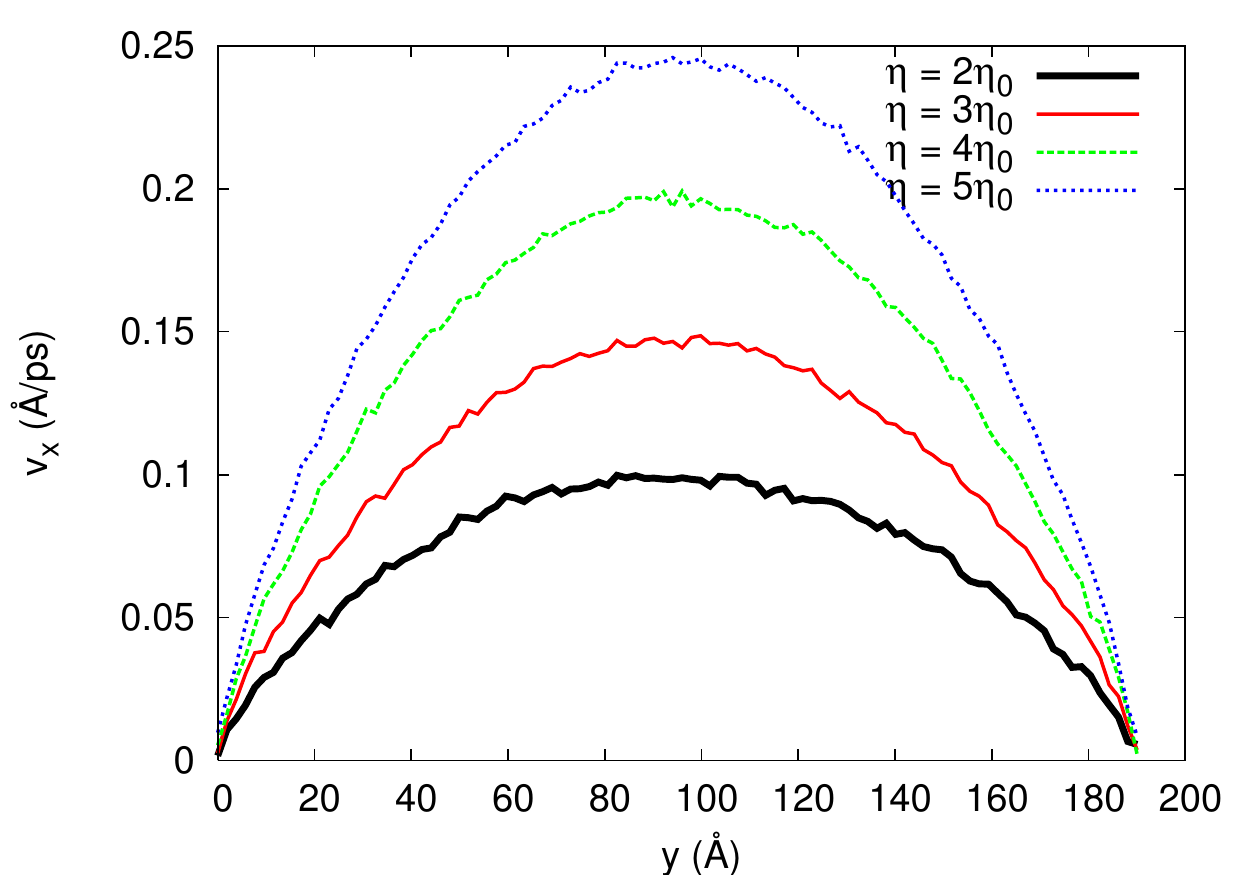}
\caption{\label{fig:flow} 
Flow velocity profiles.  For each flow rate examined here, we plot the average velocity of solvent molecules in the $x$-direction as a function of $y$.  These were obtained without the polymer.  The profiles are parabolic, due to the presence of reflective walls at $y=0$ and $y=L_y$.
}
\end{center}
\end{figure}

The flow is introduced by accelerating each
solvent particle that is not in the $y=0$ or $y=L_y$ boxes in the $x$ direction
after every rotation step according to $v_x^i \rightarrow v_x^i + \eta \Delta
t$, where $\eta$ is an acceleration parameter.
The $\eta$ values used here range from $2 \eta_0$ to $5 \eta_0$, where $\eta_0 = 625$
{\AA}/fs$^2$.  Figure \ref{fig:flow} shows average flow profiles, obtained
without the polymer.  
The P\'{e}clet number is the ratio of advective motion to thermal diffusive motion, given by
\begin{equation}
\text{Pe} = \frac{L \bar{v}_x}{D}
\end{equation}
where $L = 8$ {\AA} is the characteristic length, $\bar{v}_x$ is the average velocity 
of the solvent in the $x$ direction, and $D$ is the self-diffusion constant of a 
single residue calculated in zero flow.
Here, Pe ranges from $1.4 \times 10^{-2}$ to $3.9 \times 10^{-2}$, indicating that 
at all values of $\eta$ we examine, thermal motion is much stronger than 
advective motion
(i.e., $\text{Pe} \ll 1$). 
Prior to the start of the umbrella
sampling simulation, the solvent was equilibrated without the polymer until the
flow profiles converged; this required $40$ ns, which corresponds to roughly 
$6700$ streaming steps.

\subsection{Order parameter}
\label{op}

\begin{figure}
\begin{center}
\includegraphics[width=3.375 in]{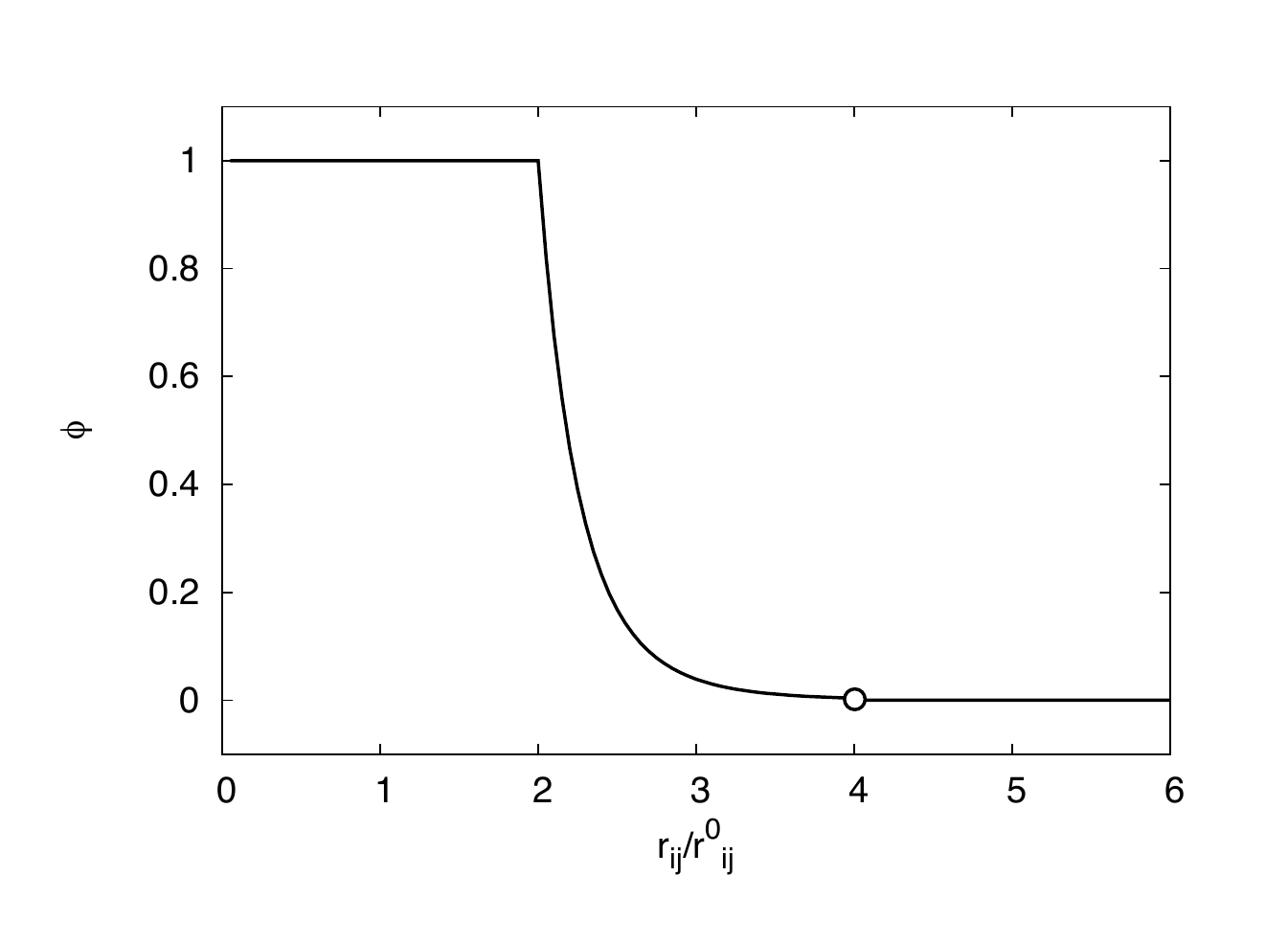}
\caption{\label{fig:phi} 
The function $\phi(r_{ij})$ that is used to calculate the order parameter $N_c$.
}
\end{center}
\end{figure}

The order parameter that we use here to distinguish between the folded and
unfolded states is an estimate of the number of native contacts that are 
made in a given configuration:
\begin{equation}
\label{eq:Nc}
N_c(t) = \sum_{i,j} \Delta_{ij}\phi(r_{ij}(t))
\end{equation} 
where $r_{ij}(t)$ is the distance between the two residues at time $t$, $\phi(r_{ij})$ is a
function that is equal to $1$ when the contact is satisfied ($r_{ij} < a_f
r^0_{ij}$), $0$ when the contact is not satisfied ($r_{ij} > 2a_f r^0_{ij}$),
and varies between $0$ and $1$ for intermediate values as $(a_f
r^0_{ij}/r_{ij})^8$, where the exponent was chosen to make the jump at $r_{ij}
= 2 a_f r^0_{ij}$ small, while being efficient to compute.  The constant $a_f =
2.0$ was used here; we found that it provided a good balance between limiting
sensitivity to fluctuations within stable states (large $a_f$) and detecting
early unfolding activity (small $a_f$).  A plot of $\phi(r_{ij})$ is shown in 
Figure \ref{fig:phi}.

We use this order parameter to define ``folded'' and ``unfolded'' basins as
$N_c \geq N_{\text{fold}}$ and $N_c \leq N_{\text{unfold}}$ respectively.  We
separate the transition path ensemble into two subensembles:  the unfolding
ensemble and the refolding ensemble.  By tracing the dynamics of each copy back
through the saved entry points, we can classify each copy as being on a trajectory
that originates in either the folded or unfolded basin.  The unfolding ensemble
is composed of all trajectories that originate in the folded basin (regardless
of whether they reach the unfolded basin or return to the folded one), and the
refolding ensemble is composed of all trajectories that originate in the unfolded basin
(regardless of whether they reach the folded basin or return to the unfolded
one).  In  other words, the ensembles are defined by the histories rather than
futures of walkers.  Each ensemble has its own set of regions that span the order parameter
space.  As shown in Dickson \textit{et al.} \cite{Dickson2009_rate}, the two
sets of regions can be seen as a single set of nonoverlapping regions in an
extended space, and transition rates between the basins can be obtained by
calculating fluxes in this extended space.

\subsection{Simulation details}

In the simulations presented here, the saved entry point lists for each region
are divided into two lists of 250 points each.  One list is dedicated to points
coming from the right (higher $N_c$) and the other to points coming from the
left (lower $N_c$).  This helps ensure that the left and right ensembles are
both well described.  An element of a list consists of the positions and
velocities of all the residues of the molecule, as well as forces from the 
previous step of the Velocity-Verlet algorithm.  Along with these data we store the
weight of the trajectory, and a time counter that is used to determine when to
perform solvent streaming steps.  We found it unnecessary to store the
coordinates of the solvent along with the flux input point, since the solvent
relaxes almost instantaneously to the presence of the polymer (data not shown),
as there are no steric interactions between the polymer and solvent.

In the work below, a cycle constitutes 2000 RNA time steps in each active sampling
region.  We allow 3000 cycles for progressive
initialization (Phase II), and another 3000 cycles with global weight updates (Phase III).  
We perform a global weight update at the beginning of Phase III, and again
every 600 cycles after that.
As will be discussed below, the number
of sampling regions used depends on the pathway observed, and is either 40 or
84 in each direction, for a total of either 80 or 168 regions in the extended
space.  The total number of sampling steps depends on how fast regions are
initialized in Phase II, but it is less than $9.6 \times 10^8$ in the 40 region
case and less than $2.02 \times 10^9$ in the 84 region case.

\section{Results}
\label{sec:results}

The RNA-under-flow system was examined at four different flow accelerations:
$\eta = 2\eta_0$, $3\eta_0$, $4\eta_0$ and $5\eta_0$.  These correspond to
P\'{e}clet numbers of $1.5 \times 10^{-2}$, $2.3 \times 10^{-2}$, $3.0 \times
10^{-2}$ and $3.9 \times 10^{-2}$, respectively.   These numbers indicate that
thermal motion is much more important than advective motion (i.e., $\text{Pe} \ll 1$),
but, as we show, there are significant flow effects.  We also examine the
equilibrium case:  $\eta = 0$.  For each flow rate we obtained folding and
unfolding rates, probability distributions for the numbers of native contacts, 
and a set of input structures to each umbrella sampling region, from which we 
can reconstruct folding and unfolding pathways.  As detailed below, the folded and unfolded 
basins were defined by our measure of the number of native contacts, $N_c$
(Section \ref{op}).

\subsection{Competing unfolding pathways}

Interestingly, we found two competing reaction pathways for the molecule.  One
pathway (``path M'') occurred by breaking contacts in the middle of the
molecule, in and around the P1 loop (residues 150-190, see Figure
\ref{fig:sec}), while the other (``path E'') occurred by breaking contacts in
and around the P5 region (residues 1-5 and 234-238), which is near the tethered
end.  We obtained pathways in duplicate for each value of $\eta$, and found a
dependence of the pathway on the flow pressure.  For $\eta \leq 3\eta_0$ we
observed path M in both trials, for $\eta = 5\eta_0$ we observed path E in both
trials, and for $\eta = 4\eta_0$ we observed path M and path E each once, which
suggests that path E is more probable for higher flow rates, and that $\eta =
4\eta_0$ is close to a transition point where the relative probabilities of the
two pathways cross over.

The folded basin for both pathways was located at $N_c \geq 960$, and the
unfolded basin was placed at the first metastable unfolded structure we
encountered along each unfolding pathway.  Although these structures could be
intermediates to further unfolded states, we will call these structures
``unfolded'', and their corresponding basins ``unfolded basins''.  For path M,
we set the unfolded basin to $N_c \leq 900$, and for path E we set the unfolded
basin to $N_c \leq 834$.  In both pathways, we define the regions in $N_c$ with
an even spacing of $\Delta N_c = 1.5$, giving us 40 regions for the unfolding
pathway in path M, and 84 regions for the unfolding pathway in path E.  There
are an equal number of regions in the refolding pathways in both cases, giving
us a total of 80 and 164 regions in paths M and E, respectively.

\subsection{Pathway analysis}

\begin{figure}
\begin{center}
\begin{tabular}{cc}
\includegraphics[width=3.375 in]{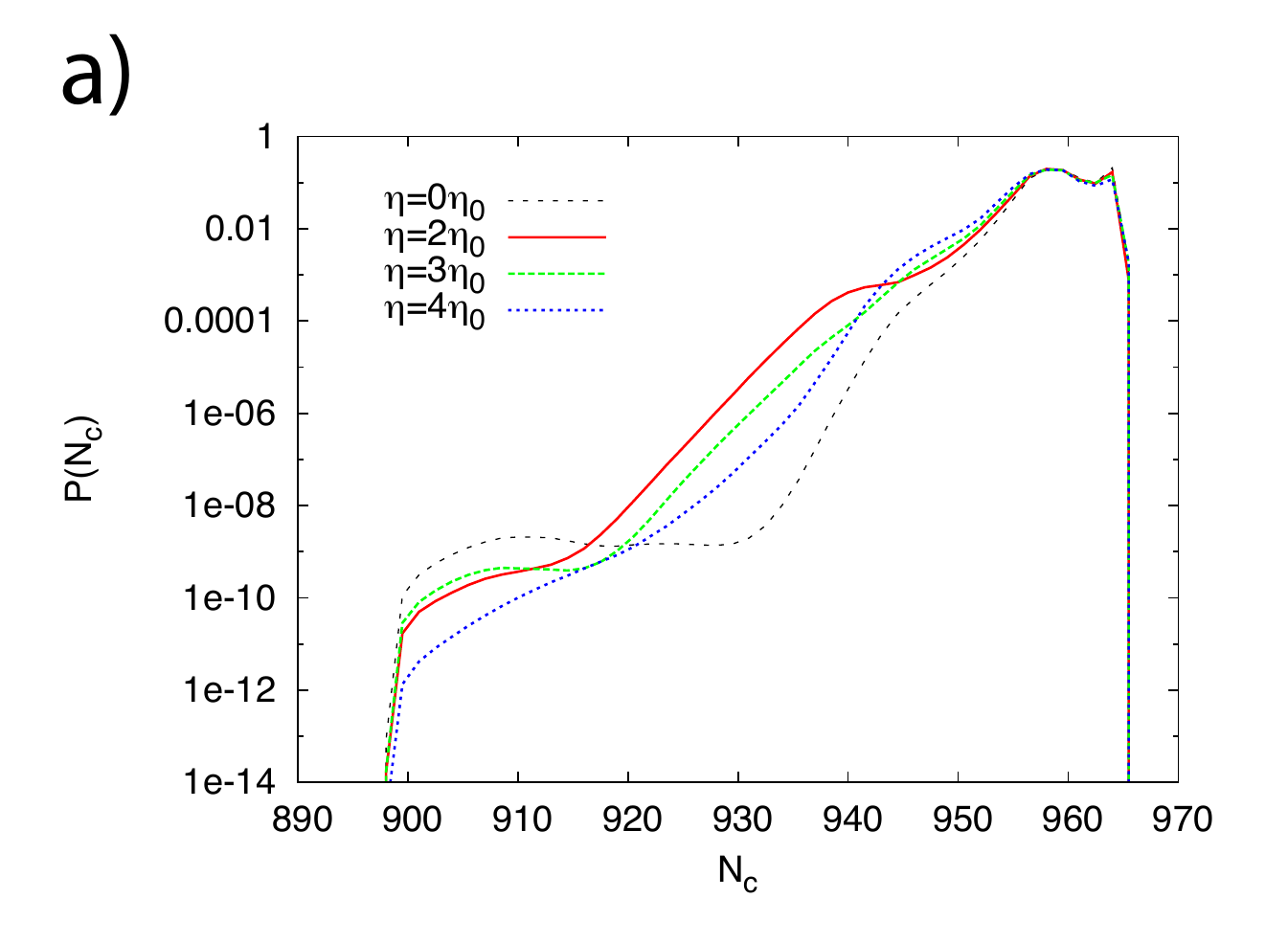} &
\includegraphics[width=3.375 in]{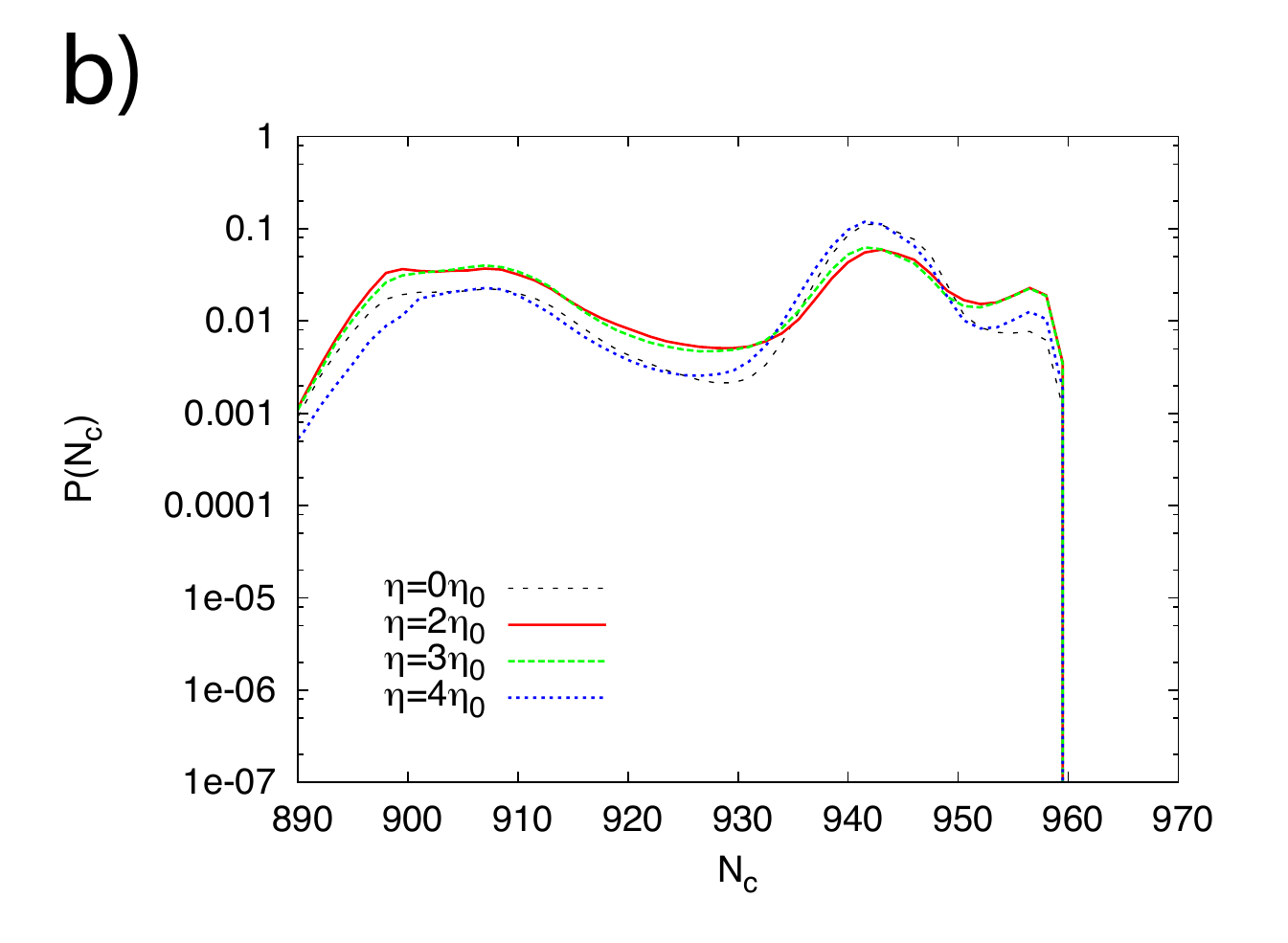} \\
\includegraphics[width=3.375 in]{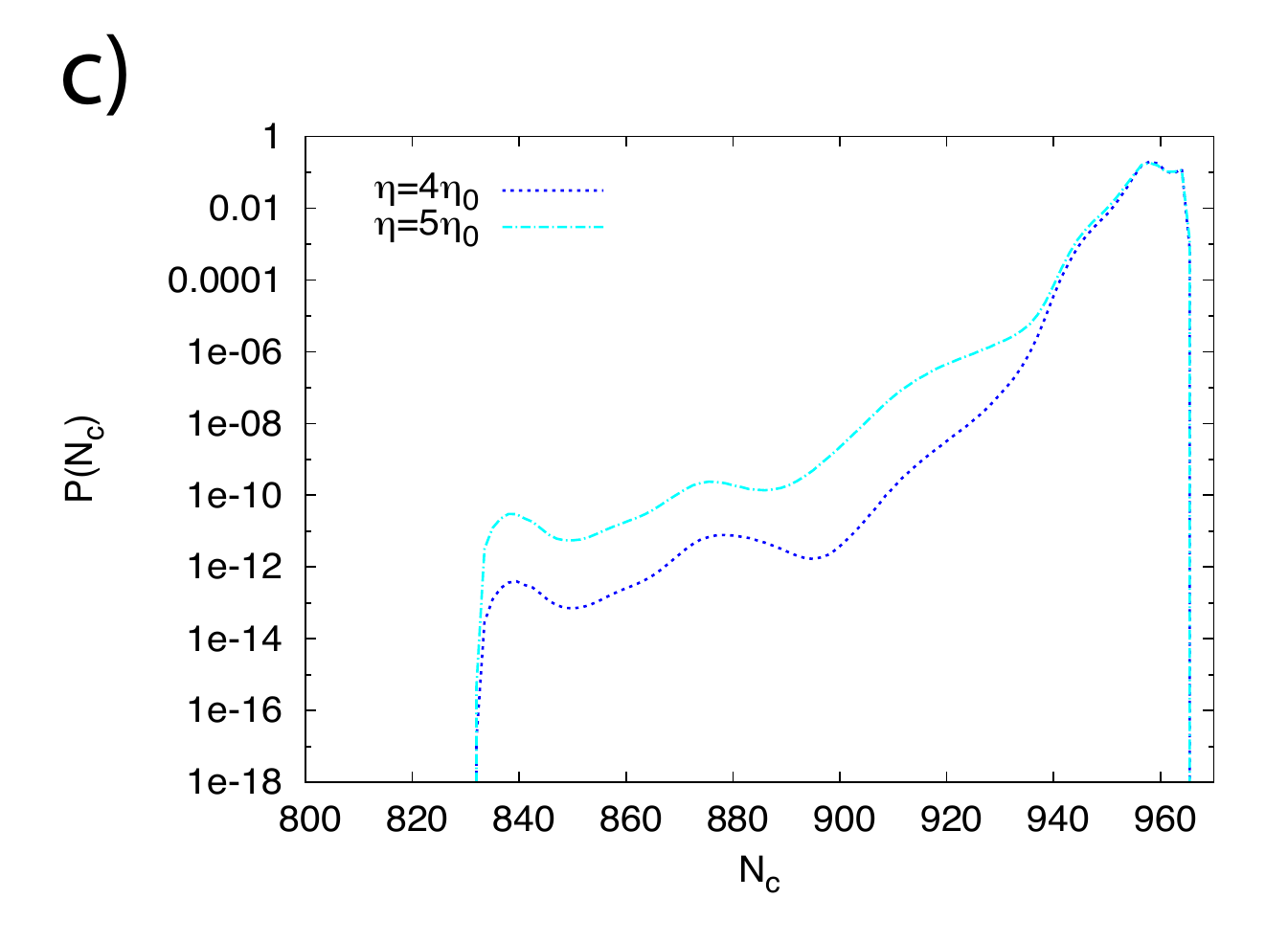} &
\includegraphics[width=3.375 in]{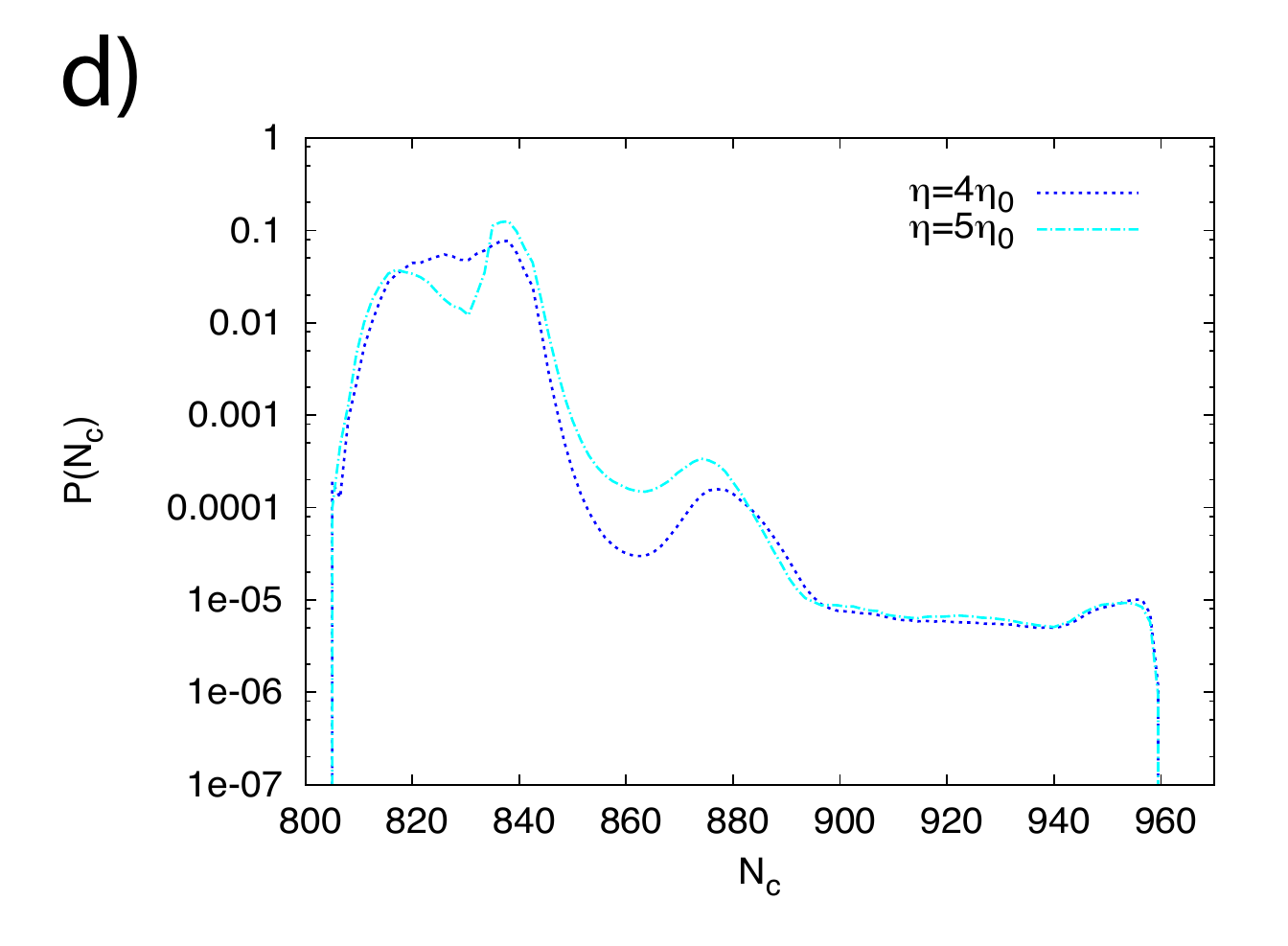} 
\end{tabular}
\caption{\label{fig:Nc} 
$N_c$-histograms for both pathways.
(a)-(b)  The unfolding and refolding ensembles of path M, respectively.  Flow pressures $\eta/\eta_0 = 0$, $2$, $3$ and $4$ are shown to be very similar.
(c)-(d)  The unfolding and refolding ensembles of path E, respectively.  Flow pressures $\eta /\eta_0= 4$ and $5$ are shown.
From (c) it is clear that this pathway is much more probable for $\eta/\eta_0 = 5$ than for $\eta/\eta_0 = 4$, and occurs approximately $100$ times faster.
}
\end{center}
\end{figure}

Probability distribution functions of the order parameter $N_c$ are shown in Figure \ref{fig:Nc}, for
both pathways, and for both the unfolding and refolding ensembles.  For path M,
we show histograms for $\eta = 2\eta_0$, $3\eta_0$ and $4\eta_0$.  In the
unfolding ensemble (Figure \ref{fig:Nc}a), there is a strong peak at $N_c =
960$ for all flow rates, corresponding to the native state.  In the refolding
ensemble (Figure \ref{fig:Nc}b), there is a peak at $N_c = 907$ corresponding
to the first metastable unfolded state, and an intermediate unfolded state at
$N_c = 942$.  For path E, we show histograms for $\eta = 4\eta_0$ and
$5\eta_0$.  Here, the refolding ensemble (Figure \ref{fig:Nc}d) shows that
there are two metastable states near the unfolded basin with peaks at $N_c =
820$ and $N_c = 838$, as well as an intermediate at $N_c = 875$.

\begin{figure}
\begin{center}
\includegraphics[width=5 in]{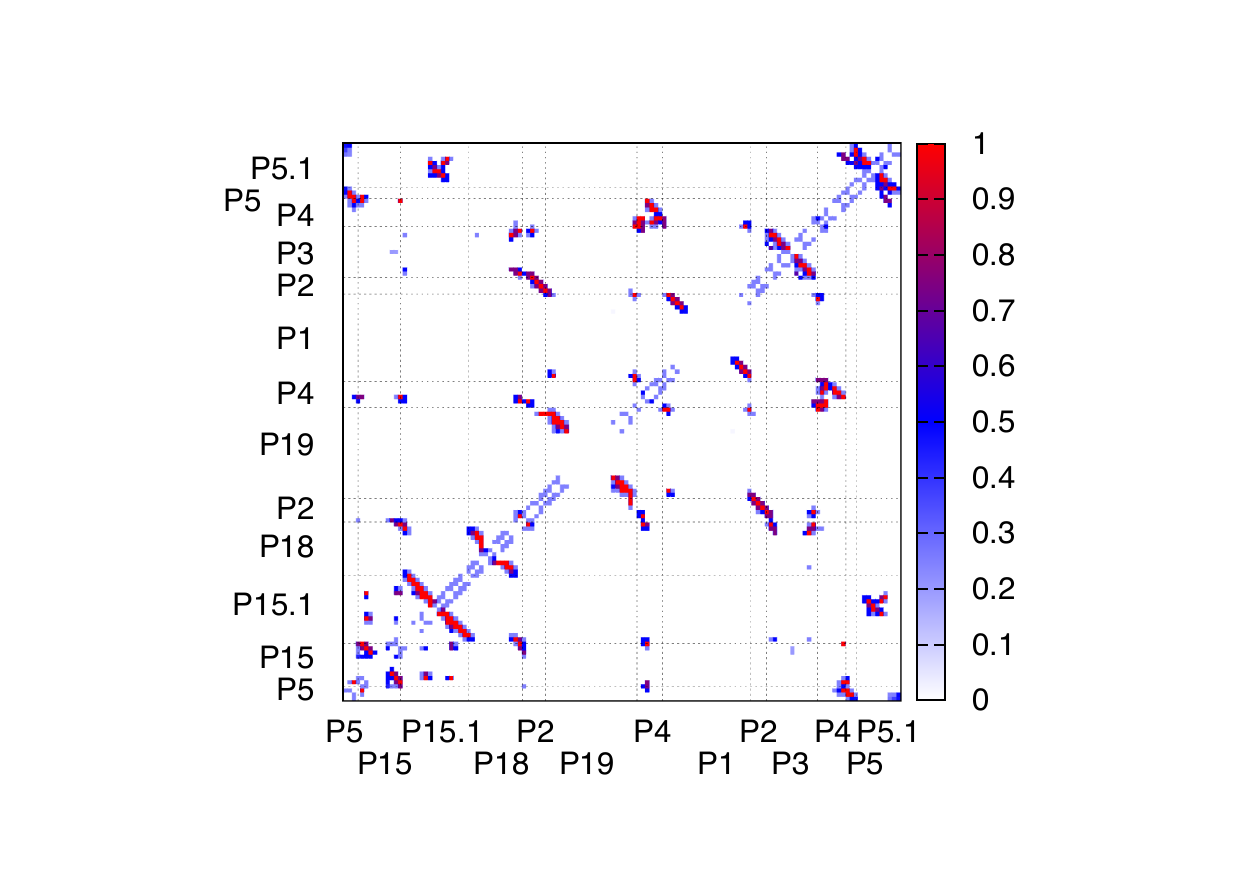}
\caption{\label{fig:contact} 
The contact map that is representative of the folded state for all flow rates examined here.  
This map was obtained using structures from entry points for the region in the unfolding ensemble with the highest value of $N_c$, at the end of a $\eta=2\eta_0$ simulation.
}
\end{center}
\end{figure}

\begin{figure}
\begin{center}
\includegraphics[width=6 in]{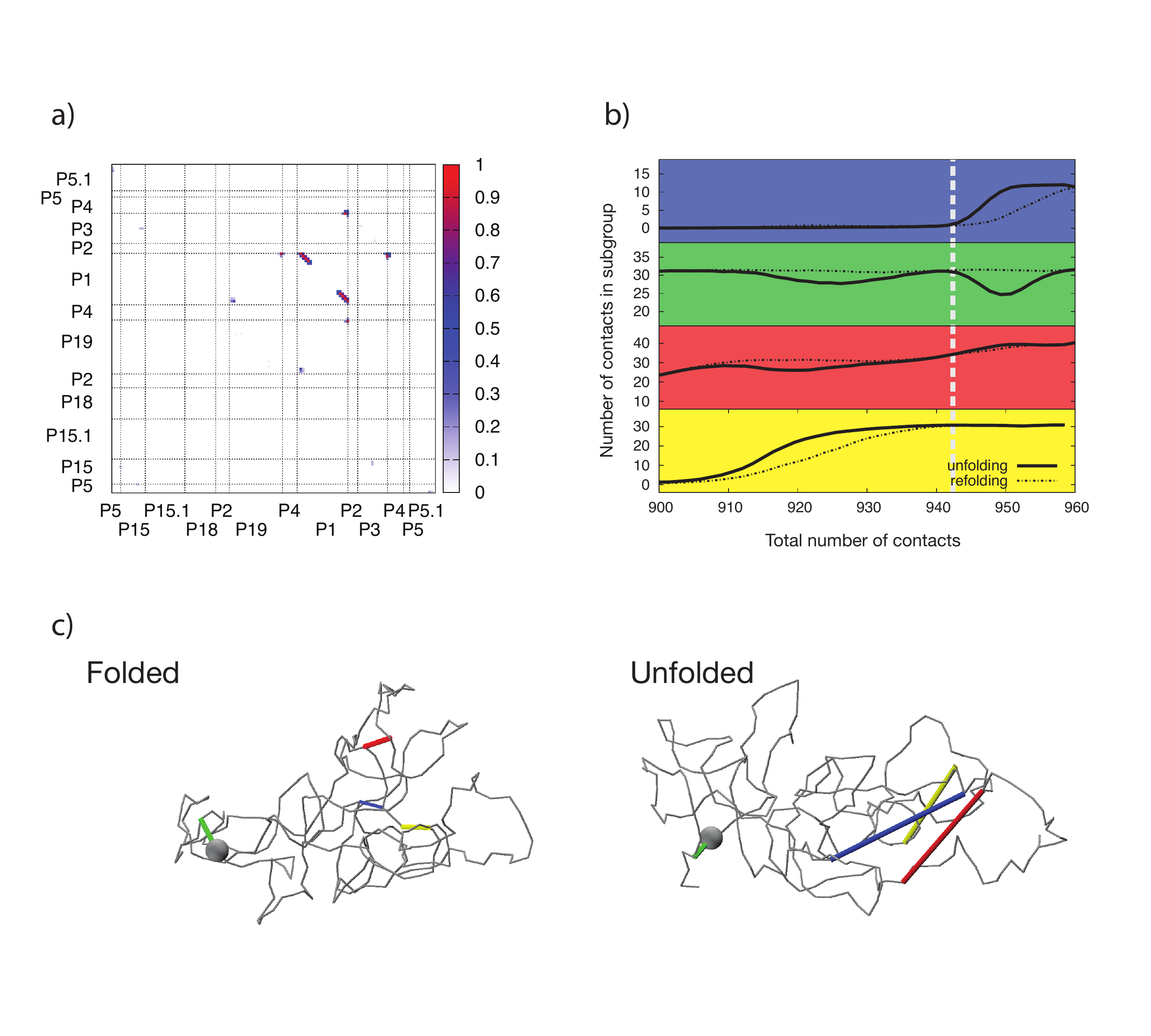}
\caption{\label{fig:sum1} 
Analysis of path M.  (a) Contact difference map obtained by subtracting the
contact map of the unfolded state from the contact map of the folded state.
This reveals the contacts which are broken along the pathway.  The colored
circles show the division of these contacts into subgroups. (b) The number of
contacts in each subgroup is plotted as a function of the total number of
contacts averaged over the $\eta = 2\eta_0$ ensemble of structures.  These
curves are computed using structures in the saved entry point lists for
every region in both the unfolding and refolding ensembles, at many times throughout Phase III of 
the simulation.  The vertical lines
show the metastable states along the refolding pathway.  (c) Representative
contacts from each group are shown on the RNA molecule for the folded and
unfolded states.
}
\end{center}
\end{figure}

\begin{figure}
\begin{center}
\includegraphics[width=6 in]{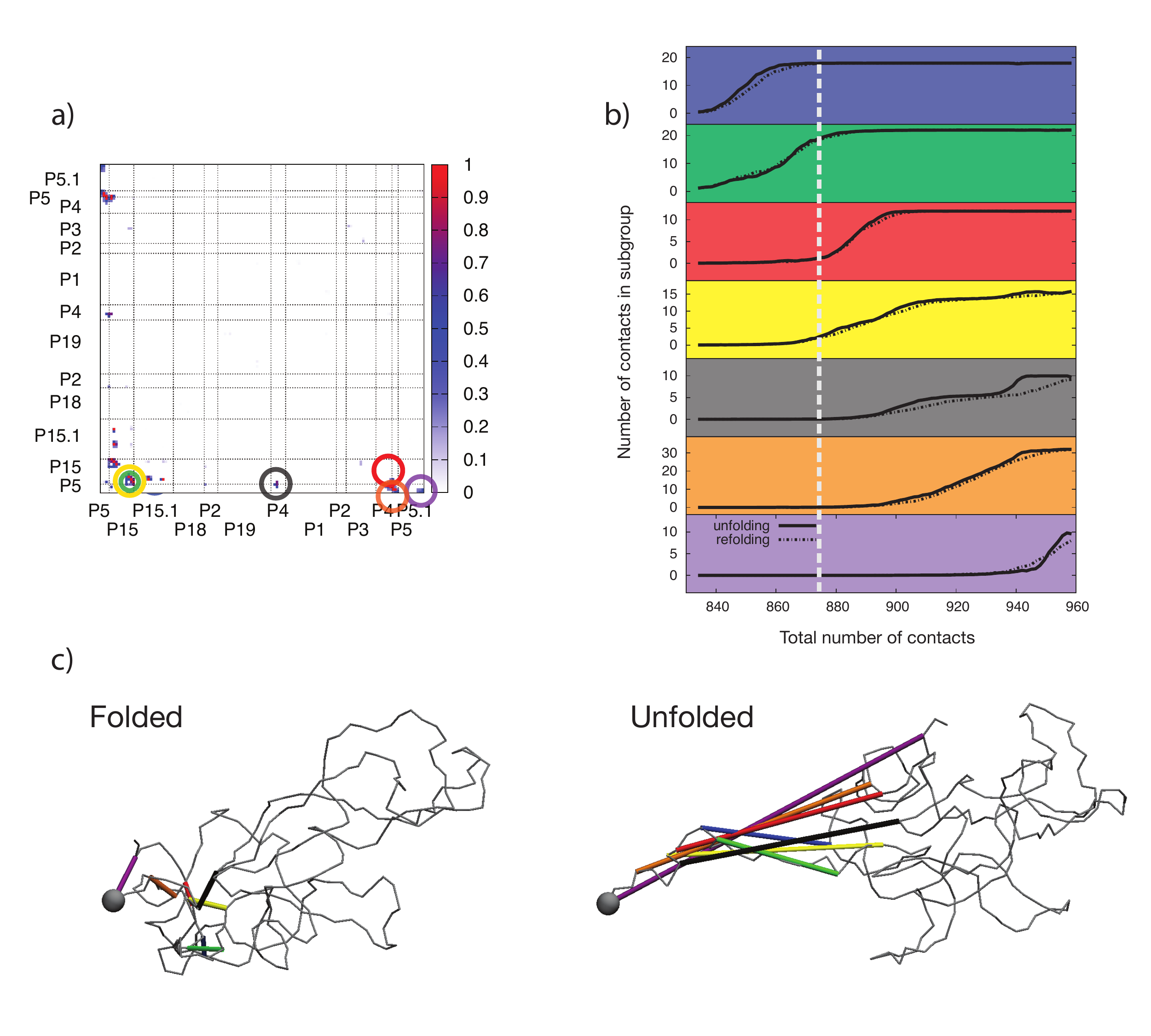}
\caption{\label{fig:sum2} 
Figure \ref{fig:sum2}.
Analysis of path E.  See descriptions of panels in Figure \ref{fig:sum1}.  (a) Contact difference map.  The green and yellow circles define subgroups of secondary and tertiary contacts within the region, respectively.
(b) Note in this pathway that there is no hysteresis between the unfolding and refolding pathways. 
}
\end{center}
\end{figure}

To characterize structures along the pathways, we construct contact difference
maps by subtracting average contact maps for the unfolded states from that for
the folded state shown in Figure \ref{fig:contact}.  The contact maps for the
unfolded states are computed using the structures from the saved entry point 
lists for the regions in the refolding ensembles with the lowest values
of $N_c$, and similarly a contact map for the folded state is computed using
structures from the saved entry point list for the region in the
unfolding ensemble with the highest value of $N_c$.  The contact difference
maps are shown in Figures \ref{fig:sum1}a and \ref{fig:sum2}a along with
characteristic structures of the folded and unfolded states (Figures
\ref{fig:sum1}c and \ref{fig:sum2}c).   Based on their kinetic behavior, we
divide the contacts into groups and track the population of each group as a
function of $N_c$ (Figures \ref{fig:sum1}b and \ref{fig:sum2}b).

The vertical lines in Figures \ref{fig:sum1}b and \ref{fig:sum2}b show the
metastable states along the refolding pathway.  For path M the local maximum
at $N_c = 942$ is associated
with the reformation of contacts in the P1 loop (subgroup 3).  For path E the
local maximum at $N_c = 875$ is associated
with the reformation of contacts in the P15 loop (subgroups 2 and 4).  For path
M we observe that the unfolding and refolding ensembles do not overlap.
Specifically, contacts between the endpoints of the molecule (P5-P5.1 contacts)
break and reform along the unfolding pathway, but remain intact during the
refolding pathway.  In this regard, it is important to keep in mind that the
unfolding ensemble, as defined in Section \ref{op} contains both
folded-to-unfolded trajectories and folded-to-folded trajectories.  The fact
that the feature in question appears in analogous calculations for the
reversible system ($\eta=0$), where there can be no hysteresis, suggests that
the P5-P5.1 contacts are broken along folded-to-folded trajectories, and that
this process is not a causal part of the path M unfolding mechanism.

\subsection{Transition rates}

\begin{figure}
\begin{center}
\includegraphics[width=3.375 in]{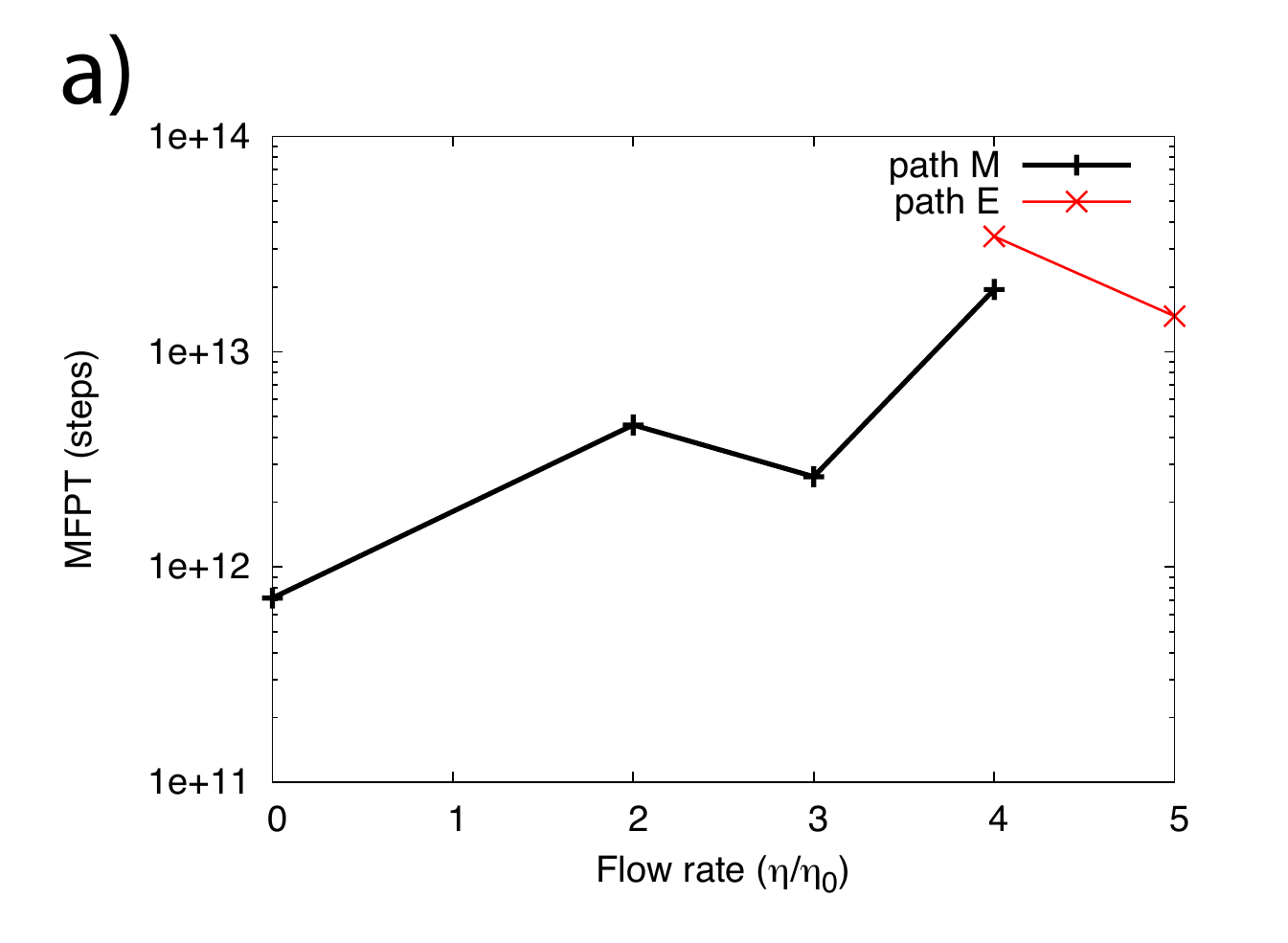}
\includegraphics[width=3.375 in]{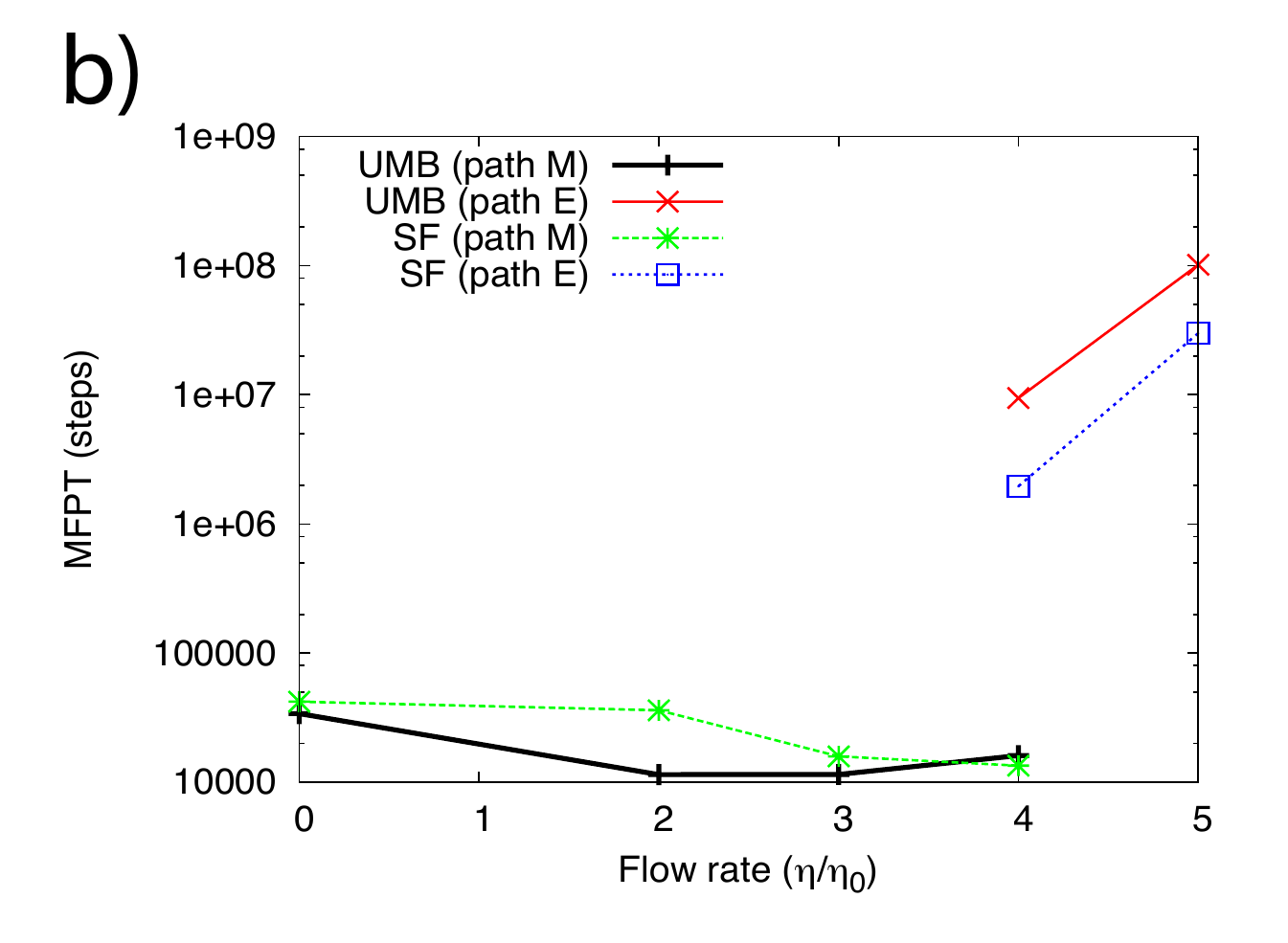}
\caption{\label{fig:mfpt} 
Figure \ref{fig:mfpt}.
(a) Mean first passage times for unfolding events, as predicted by NEUS.
For path M, this is the average number of steps required to go from $N_c = 960$ to $N_c = 900$, and for path E, this is the average number of steps to go from $N_c = 960$ to $N_c = 834$.
(b) Mean first passage times for refolding events, comparing umbrella sampling (UMB) and straightforward trajectories (SF).  These agree to within an order of magnitude.
}
\end{center}
\end{figure}

The mean first passage times of the unfolding processes are given in Tables
\ref{tab:mfpt1} and \ref{tab:mfpt2} for paths M and E respectively. 
These range from $6.8$ to $460$ ms for path M and $590$ to $1400$ ms for path E.
As each dynamics step is $40$ fs, these correspond to numbers of dynamics steps
between $1.7 \times 10^{11}$ and $3.4 \times 10^{13}$.
The unfolding and refolding MFPTs are shown as functions of flow pressure in
Figure \ref{fig:mfpt}.

\begin{table} 
\caption{\label{tab:mfpt1}  
Unfolding and refolding mean first passage times for path M, obtained for $\eta = 0$, $2\eta_0$, $3\eta_0$ and $4\eta_0$.  
For refolding pathways, the MFPTs from umbrella sampling (NEUS) and straightforward sampling (SF) are shown.
}
\begin{ruledtabular}
\begin{tabular}{cccc}
$\eta/\eta_0$&Unfolding (NEUS) (in ms)&Refolding (NEUS) (in ns)&Refolding (SF) (in ns) \\ \hline
$0$&$220$&$1.5$&$1.7$ \\
$2$&$6.8$&$0.82$&$1.4$ \\
$3$&$110$&$0.85$&$0.6$ \\
$4$&$460$&$0.91$&$0.5$ \\
\end{tabular}
\end{ruledtabular}
\end{table}

\begin{table} 
\caption{\label{tab:mfpt2}  
Unfolding and refolding mean first passage times for path E, obtained for $\eta = 4\eta_0$ and $5\eta_0$.  
For refolding pathways, the MFPTs from umbrella sampling (NEUS) and straightforward sampling (SF) are shown.
}
\begin{ruledtabular}
\begin{tabular}{cccc}
$\eta/\eta_0$&Unfolding (NEUS) (in ms)&Refolding (NEUS) (in $\mu$s)&Refolding (SF) (in $\mu$s) \\ \hline
$4$&$1400$&$0.38$&$0.08$ \\
$5$&$590$&$4.1$&$1.2$ \\
\end{tabular}
\end{ruledtabular}
\end{table}

The unfolding rates show counter-intuitive behavior for path M.  
The MFPT increases with the flow rate; unfolding becomes more difficult as greater flow is applied to the system.
Such behavior could be caused by larger flow gradients at the surface 
causing nucleotides in the P1 loop to be pushed together rather than pulled apart.
For path E, the MFPT for unfolding decreases with increasing flow rate, 
although we only have two data points to establish this trend.
We note that the MFPTs for path M and path E are not directly comparable, since the former 
measures the average amount of time to go from $960$ to $900$ contacts, and the latter measures 
the average amount of time to go from $960$ to $834$ contacts.

The rates of refolding are also given in
Tables \ref{tab:mfpt1} and \ref{tab:mfpt2}.  
They are much faster, which
makes comparisons with straightforward trajectories possible.  We use the
umbrella sampling saved entry point lists to generate an initial unfolded
ensemble for each flow rate, since umbrella sampling is our only access to
physically weighted unfolded states.  We compare refolding rates for both
pathways and all flow pressures, which agree to within an order of magnitude.
For path M, the refolding MFPT is short ($\sim 1$ ns), and relatively constant with varying flow rate.
For path E, the refolding MFPTs are longer, since the unfolded state is more stable, and increase with increasing flow rate:  $0.3$ $\mu$s for $\eta = 4\eta_0$, and $4.1$ $\mu$s for $\eta = 5\eta_0$.
This behavior suggests that higher flow fields stabilize the unfolded state.

To illustrate the importance of the enhanced sampling algorithm for the
unfolding simulations, we computed 16 independent trajectories of 16 $\mu$s ($4
\times 10^8$ dynamics steps) starting from structures taken from the folded basin. 
These trajectories were run using $\eta = 5\eta_0$, and ``unfolding'' was defined as 
reaching $900$ contacts instead of the usual $834$ for path E, in order to increase the probability of observing an unfolding event.
Using NEUS we found the MFPT for this process was $0.21$ ms, making the length of the straightforward trajectories
12.5\% of the predicted MFPT, and no unfolding events were
observed.  These simulations required 30 days of computation on 16 2.5 GHz
Intel Xeon processors.  
This also emphasizes the computational benefit of parallelization, as the $\sim 2
\times 10^9$ steps for the largest umbrella sampling simulations were completed
in $\sim 30$ h of computation on 64 processors.
However, even if a similar parallelization scheme using 64 processors was employed 
for straightforward trajectories, it would still take an average of 58 years of 
computer time to observe a single path E unfolding trajectory for $\eta = 4\eta_0$, 
and many times that to observe an ensemble of unfolding events.

\section{Conclusion}
\label{sec:conc}

Here we have presented a parallel version of NEUS and applied it to a
coarse-grained macromolecular system driven far from equilibrium by flow.  
We obtained folding and unfolding rates and mechanisms for a range of flow speeds.
This range was chosen to be physically reasonable yet result in significant
flow effects.  It is large compared to $1.6 \times 10^{-5}$, the P\'{e}clet number
of the flow used to change the magnesium ion concentrations in the RNase P RNA
single molecule experiments of Qu \textit{et al.} \cite{Qu2008},
and our simulations suggest that flow did not contribute to the dynamics discussed in
\cite{Qu2008,Li2009}, at least at moderately high magnesium ion concentrations,
which strongly favor the folded state.  A lack of knowledge of the structure of
the RNA at low magnesium ion concentrations prevents us from assessing that
situation.

Due to
the stability of the native state, unfolding transitions were extremely slow,
occurring as slowly as once in every $3.4 \times 10^{13}$ dynamics steps, or
every $1.4$ s in real time.  We observed two different unfolding pathways, 
one where secondary contacts were broken in the P1 loop, and another where contacts were broken in 
and around the P5 loop, which is near the tethered endpoint.
We defined unfolded and folded states using an order
parameter that measures the number of native contacts.  
If one were to use more than one order parameter, sampling could be
enforced separately along these two pathways.  This would allow for a more
precise description of the competition between the two pathways for a given
flow rate, and a description of the transition between the pathways of maximum
probability as the flow rate changes.  Work is currently underway to acheive
this goal.
The parallelization strategy presented here for piecewise sampling methods will enable
treatment of increasingly complex order parameter spaces as large-scale computational
architectures continue to grow in size.

\section*{Acknowledgments}

We would like to thank Nicholas Guttenberg and Jonathan Weare for useful discussions on the algorithm, and Glenna Smith and Norbert Scherer for help with the RNA model.
This work was supported by National Science Foundation grant No. MCB-0547854, an Argonne-University of Chicago Strategic Collaborative Initiative Award, and the Natural Sciences and Engineering Research Council.  We gratefully acknowledge the computing resources provided on ``Fusion,'' a 320-node computing cluster operated by the Laboratory Computing Resource Center at Argonne National Laboratory, which is supported by the Office of Science of the U.S. Department of Energy under contract DE-AC02-06CH11357.

\end{document}